\documentclass{ws-procs975x65}

\usepackage{graphicx}
\usepackage{euscript,amssymb}
\usepackage{epsfig}
\usepackage{color}
\usepackage{amsfonts}
\usepackage{amsmath}
\usepackage{amssymb}
\usepackage{fancyhdr}
\usepackage{esint}

\begin{document}

\title{QUANTUM FIELD THEORY ON LQC BIANCHI SPACETIMES}

\author{ANDREA DAPOR$^*$ and JERZY LEWANDOWSKI$^\dagger$}

\address{Instytut Fizyki Teoretycznej, Uniwersytet Warszawski,\\
ul. Ho\.{z}a 69, 00-681 Warsaw, Poland\\
$^*$E-mail: adapor@fuw.edu.pl\\
$^\dagger$E-mail: jerzy.lewandowski@fuw.edu.pl}

\author{YASER TAVAKOLI}
\address{Departamento de F\'{\i}sica,
Universidade da Beira Interior,\\
Rua Marqu\^{e}s d'Avila e Bolama, 6200 Covilh\~{a}, Portugal\\
E-mail: tavakoli@ubi.pt}

\begin{abstract}
Quantum theory of a scalar field is developed on the LQC Bianchi I space-time.
By comparing the the quantum field theory for a single mode on classical and quantum background geometries we find that an effective Bianchi 
I space-time emerges.  We show that by disregarding the back-reaction no Lorentz-violation is present, despite the effective metric being different 
than the classical Bianchi I one.
\end{abstract}


\bodymatter

\section{QFT on classical Bianchi I space-time}
We consider the background Bianchi I (BI) space-time manifold to be topologically $M=\mathbb{R}\times\mathbb{T}^3$, equipped with $(x_0, x^j)$. In terms of $SU(2)$ connection variables $A^{i}_{a} = c^{i} \delta^{i}_{a}$ and $E^{a}_{i} = p_{i} \delta^{a}_{i}$, the phase space variables of gravity,  $\Gamma_{\mathrm{gr}}$, are given by $p_1=a_2a_3$, $p_2=a_3a_1$ and $p_3=a_1a_2$. The background BI space-time reads
\begin{eqnarray}
g_{\mu \nu} dx^{\mu} dx^{\nu} = -N_{x_{0}}^{2}(x_{0}) dx_{0}^{2} + |p_1p_2p_3|\sum_{i = 1}^{3} \frac{(dx^{i})^{2}}{p_{i}^{2}(x_{0}) },
\label{24}
\end{eqnarray}
where $x^j\in \mathbb{T}^3$, with $x_0\in \mathbb{R}$ being a generic time coordinate. A real (inhomogeneous) scalar field $\phi(x_0,\vec{x})$ is considered on this background space-time, whose Lagrangian is $\mathcal{L}_{\phi} = \frac{1}{2} (g^{\mu \nu} \partial_{\mu} \phi \partial_{\nu} \phi - m^{2} \phi^{2})$. For the pair $(\phi,\pi_\phi)$,
the classical solutions of the equation of motion can be expanded in terms of the Fourier transformation \cite{DLT, AKL},
from which $\phi_{\vec{k}}$ and  $\pi_{\vec{k}}$ satisfy the Poisson bracket $\{\phi_{\vec{k}}, \pi_{\vec{k}'}\} = \delta_{\vec{k}, -\vec{k}'}$.

The total Hamiltonian of the system can be described as sum of the Hamiltonian of decoupled harmonic oscillators, namely $H_{\vec{k}}(x_{0})$, in terms of two conjugate variables $q_{\vec{k}}$ and $p_{\vec{k}}$ associated with each mode $\vec{k}$ of the field \cite{DLT}.
In quantum theory, for each mode  $\vec{k}$, the Hilbert space of the matter is $\mathcal{H}_{\vec{k}} = L_{2}(\mathbb{R}, dq_{\vec{k}})$, therefore, the time $x_{0}$-evolution  is generated by the time-dependent Hamiltonian operator $\widehat{H}_{\vec{k}}(x_{0})$ via Schroedinger equation \cite{DLT}
\begin{align}
& i\hbar \partial_{x_{0}} \psi(x_{0},q_{\vec{k}}) = \widehat{H}_{\vec{k}}(x_{0}) \psi(x_{0}, q_{\vec{k}}) \notag \\
& \ \ = \frac{N_{x_{0}}(x_{0})}{2 \sqrt{|p_{1}(x_{0}) p_{2}(x_{0}) p_{3}(x_{0})|}}
\left[\widehat{p}_{\vec{k}}^2 + \left(\sum_{i = 1}^{3} (p_{i} k_{i})^{2} + |p_{1} p_{2} p_{3}| m^{2}\right) \widehat{q}_{\vec{k}}^{2}\right] \psi(x_{0}, q_{\vec{k}})
\label{classfield}
\end{align}

\section{QFT on quantum Bianchi I space-time}
%
In LQC of BI model, Gauss and vector constraints are automatically satisfied and we are left only with the (homogeneous part of) scalar constraint, given by the gravity part and the scalar field part, $C_{\text{gr}}$ and $C_T$: the total Hamiltonian reads
$C_{\text{geo}} = C_{\text{gr}} + C_{T}$.
The massless scalar field $T$ plays the role of \emph{relational time} in LQC \cite{awe2}. Therefore, it is convenient to work in terms of the coordinate $(T, x^{i})$, by considering  that $N_{x_0}dx_0=N_T dT$, where $N_{T} =  \sqrt{|p_{1} p_{2} p_{3}|}/P_{T}$ in Eq. (\ref{24}).

The kinematical Hilbert space of BI model, ${\cal H}_{\text{kin}}$, is given as  $\mathcal{H}_{\text{gr}}\otimes\mathcal{H}_{\text{T}}$, with $\mathcal{H}_{\text{gr}}$ being the Hilbert space of the gravitational sector spanned by the $\hat{p}_i$-eigenstates $|\vec{\lambda}\rangle := |\lambda_1, \lambda_3, \lambda_3\rangle$, and $\mathcal{H}_{\text{T}} = L_{2}(\mathbb{R}, dT)$ is the Hilbert space of the scalar field which is quantized according to Schroedinger picture. Therefore, the scalar constraint operator $\widehat{C}_{\text{geo}}$ is well-defined on $\mathcal{H}_{\text{kin}}$:
$\widehat{C}_{\text{geo}} = -\frac{1}{2} (\hbar^{2}\partial_{T}^{2} \otimes \mathbb{I}) - \frac{1}{2} (\mathbb{I} \otimes \Theta)$.
Physical states  $\Psi_o(T, \vec{\lambda}) \in \mathcal{H}_{\text{kin}}$ are the (positive frequency) solutions to \cite{awe2}
\begin{align}
-i\hbar \partial_{T} \Psi_o(T, \vec{\lambda})  = \sqrt{|\Theta|} \Psi_o(T, \vec{\lambda}) =: \widehat{H}_o\Psi_o(T, \vec{\lambda}) \  .
\label{23-a}
\end{align}

In the presence of a real scalar field on the BI background quantum geometry, the kinematical Hilbert space for each single mode $\vec{k}$ becomes
${\cal H}_{\text{kin}}^{(\vec{k})} = {\cal H}_{\text{geo}} \otimes L^2(\mathbb{R}, dq_{\vec{k}})$.
So that, the total scalar constraint of the system 
reads
$\widehat{C}_{\tau, \vec{k}} = -\frac{\hbar^{2}}{2} (\partial_{T}^{2} \otimes \mathbb{I}_{\text{gr}} \otimes \mathbb{I}_{\vec{k}})
- \frac{1}{2}(\mathbb{I}_{T} \otimes \Theta \otimes \mathbb{I}_{\vec{k}}) + (\mathbb{I}_{T} \otimes \widehat{H}_{\tau, \vec{k}})$,
where
\begin{eqnarray}
\widehat{H}_{\tau, \vec{k}} = \frac{1}{2} \left[\widehat{p}_{\vec{k}}^{2} + \left(\sum_{i = 1}^{3} \widehat{p}_{i}^{2} k_{i}^{2} + |\widehat{p}_{1} \widehat{p}_{2} \widehat{p}_{3}| m^{2}\right) \widehat{q}_{\vec{k}}^{2}\right].
\end{eqnarray}
Then, physical states, $\Psi(T,\vec{\lambda}, q_{\vec{k}})$, on ${\cal H}_{\text{phys}}^{(\vec{k})}=Ker~(\widehat{C}_{\tau, \vec{k}})$ are the solutions to:
\begin{align}
-i \hbar \partial_{T} \Psi(T, \vec{\lambda}, q_{\vec{k}}) =  \left[\widehat{H}_{o} - \widehat{H}_{o}^{-\frac{1}{2}} \widehat{H}_{\tau, \vec{k}} \widehat{H}_{o}^{-\frac{1}{2}}\right] \Psi(T, \vec{\lambda}, q_{\vec{k}}).
\label{QFTQG}
\end{align}
Notice that, here we have used the test field approximation where the backreaction of the scalar field on geometry was disregarded.

\section{Effective geometry and Lorentz symmetry}

Let us take the classical limit for the geometrical degrees of freedom through the interaction picture. The physical state of the system in this case reads: $ \Psi(T, \vec{\lambda}, q_{\vec{k}}) = \Psi_o(T, \vec{\lambda}) \otimes \psi(T, q_{\vec{k}})$,
where the geometry evolves through $\widehat{H}_o$, i.e., $-i\hbar \partial_T \Psi_o = \widehat{H}_o \Psi_o$:
Then, by substituting this in Eq. (\ref{QFTQG}) we find:
\begin{align}
i \hbar \partial_{T} \psi(T, q_{\vec{k}}) & = \frac{1}{2} \left[\langle \widehat{H}_{o}^{-1} \rangle\widehat{p}_{\vec{k}}^{2}
+ \langle \widehat{H}_{o}^{-\frac{1}{2}} \left(\sum_{i = 1}^{3} \widehat{p}_{i}^2(T) k_{i}^{2} \right. \right. \notag \\
&\ \  \ \ \ \ \ \left. \left.  + |\widehat{p}_{1}(T) \widehat{p}_{2}(T) \widehat{p}_{3}(T)| m^{2}\right) \widehat{H}_{o}^{-\frac{1}{2}} \rangle \widehat{q}_{\vec{k}}^{2}\right] \psi(T, q_{\vec{k}}),
\label{58}
\end{align}
where $\langle \widehat{A}(T) \rangle$ denotes the expectation value on $\Psi_{o}(0, \vec{\lambda})$ of the gravitational operator $\widehat{A}(T) = e^{-iT \widehat{H}_o/\hbar} \widehat{A} e^{iT \widehat{H}_o/\hbar}$. By setting $x_0=T$ in Eq. (\ref{classfield}) and by comparing with Eq. (\ref{58}), we find a Shroedinger equation associated an \emph{effective BI metric},
whose components $\bar{N}$ and $\bar{p}_i$ (for a massless scalar field, $m=0$) are
\begin{align}
\bar{N}(T) =  \left(\langle \widehat{H}_{o}^{-1} \rangle \prod_{i = 1}^{3} \langle \widehat{H}_{o}^{-\frac{1}{2}} \widehat{p}_{i}^{2}(T) \widehat{H}_{o}^{-\frac{1}{2}} \rangle\right)^{\frac{1}{4}}, \ \ \
\bar{p}_{i} = \left[\dfrac{\langle \widehat{H}_{o}^{-1/2} \widehat{p}_{i}^{2}(T) \widehat{H}_{o}^{-1/2} \rangle}{\langle \widehat{H}_{o}^{-1} \rangle}\right]^{\frac{1}{2}}.
\label{61-b}
\end{align}
Therefore, the effective BI space-time is emerged in terms of the expectation values of the gravitational operators on
the quantum geometry state $\Psi_{o}$, whose components do not depend on modes $\vec{k}$.

A prediction of many approaches to quantum gravity comes from the study of \emph{in vacuo} ``dispersion relation" (i.e., the relation between the frequency and the wave-vector of a mode of a field). The wave equation on the effective geometry, described by Eq. (\ref{61-b}), can be obtained as \cite{DLT}
\begin{eqnarray}
\frac{d^{2}\mathcal{Q}_{\vec{k}}}{dT^{2}} + \Omega_{T,\vec{k}}^{2} \mathcal{Q}_{\vec{k}} = 0,
\label{WaveEq-Eff}
\end{eqnarray}
where $\mathcal{Q}_{\vec{k}}$ denotes the modified modes $q_{\vec{k}}$ given by $\mathcal{Q}_{\vec{k}} := q_{\vec{k}}/\sqrt{\langle \widehat{H}_{o}^{-1}\rangle}$, and
\begin{align}
\Omega_{T,\vec{k}}^2(T) = \langle \widehat{H}_{o}^{-1}\rangle\sum_i k_{i}^{2}\langle \widehat{H}_{o}^{-\frac{1}{2}} \widehat{p}_{i}^{2}(T) \widehat{H}_{o}^{-\frac{1}{2}} \rangle,
\end{align}
is the (modified) dispersion relation of the test field on the effective geometry.
The 3-velocity of modes for the massless scalar field propagating on this effective geometry is given by
$\|V\|^{2}  = -\sum_{i} \frac{\bar{g}_{ii}}{\bar{g}_{00}} \left(\frac{d \Omega_{T,\vec{k}}}{dk_{i}}\right)^{2} = 1 $, which
indicates that no Lorentz-violation is present in our model.

To summarize, we have developed the QFT of single modes on the BI LQC space-time.
By disregarding the back-reaction we obtained an ``effective geometry" felt by quanta of matter.
It was shown that no Lorentz-violation in BI space-time is present at 0th order (test field approximation).

\section*{Acknowledgments}

This work was partially supported by the grant 182/N-QGG/2008/0 (PMN) of Polish Ministerstwo Nauki i Szkolnictwa Wy\.{z}szego. YT thanks the ESF sponsored network `Quantum Geometry and Quantum Gravity' for a short visit grant. He is supported by FCT (Portugal) through the fellowship SFRH/BD/43709/2008.


\begin{thebibliography}{9}
\bibitem{DLT} A.~Dapor, J.~Lewandowski, Y.~Tavakoli, {\em Phys. Rev.} {\bf D} \textbf{86}, 064013 (2012).
\bibitem{AKL} A.~Ashtekar, W.~Kaminski, J.~Lewandowski,  {\em Phys. Rev.} {\bf D} \textbf{79}, 0644030 (2009).
\bibitem{awe2} A.~Ashtekar, E.~Wilson-Ewing, {\em Phys. Rev.} {\bf D} \textbf{79}, 083535 (2009).


\end{thebibliography}

\end{document}